\documentstyle[preprint,eqsecnum,aps]{revtex}
\begin{document}
\def \beq{\begin{equation}}
\def \eeq{\end{equation}}
\def \beqarr{\begin{eqnarray}}
\def \eeqarr{\end{eqnarray}}

\title{Bimerons in Double Layer Quantum Hall Systems}
\author{Sankalpa Ghosh and R. Rajaraman\cite{byline1}}

\address{School of Physical Sciences \\
Jawaharlal Nehru University\\ New Delhi 110067, \ INDIA\\ }

\maketitle
\begin{abstract}

In this paper we discuss bimeron pseudo spin
textures for  double layer quantum hall systems with filling factor $
\nu =1$. Bimerons are excitations corresponding to bound pairs of merons
 and anti-merons.  Bimeron solutions have already been studied at great
length by other groups by minimising the microsopic Hamiltonian
between microscopic trial wavefunctions. 
 Here we calculate them by numerically solving
coupled nonlinear partial differential equations arising from
extremisation of  the effective action for pseudospin textures.
We also calculate the different contributions to the energy of 
our bimerons , coming from pseudospin stiffness, capacitance 
and coulomb interactions between the merons. 
Apart from  augmenting earlier results , this allows us to check
how good an approximation it is to think of the bimeron
as a pair of rigid objects (merons) with logarithmically
growing energy, and with electric charge ${1 \over 2}$.
Our differential equation
approach  also allows us to study the dependence
of the spin texture as a function of the distance between
merons, and the inter layer distance. Lastly, the technical problem
of solving coupled nonlinear partial differential equations, subject to
the special boundary conditions of bimerons is interesting in its own 
right.

\end{abstract}

\section{Introduction}

The fascinating discoveries of the quantum Hall effect (QHE)
originally found in single two dimensional electron layers, have
also been extended to double layer systems , thanks to the
development of techniques for growing GaAs heterostructures
containing two separated layers of two-dimensional electron gas
(see for example references \cite{Eisen}). Apart from  finding
plateaus in Hall conductivity at total filling fractions $\nu$
corresponding to the "direct sum" of the familiar integral and
odd-denominator fractional QHE in the individual single layers,
experiments also show the occurence of new plateaus which are
intrinsic to double-layer systems and rely on interlayer quantum
coherence and correlations.  On the theoretical front, a large
body of work has already been done on double-layer systems.  An
extensive list of references to this literature has been given
in the lucid review of this subject by Girvin and MacDonald
\cite{GirvMac} and in the paper by Moon ${\it et al}$ \cite{Moon}.
 
Generally one  analyses double layer systems by attributing to the
electrons,  in addition to their spatial coordinates on the
plane,  a two-component "pseudospin" whose up and down
components refer to the amplitude for the electron to be in the
first and second layers respectively. The real physical spin of
the electrons is assumed , as a staring approximation, to be
fully polarised by the strong magnetic field and hence frozen 
as a degree of freedom.  However, even when real physical spin is
suppressed, the use of a pseudospin to represent  the layer
degree of freedom maps the double layer spinless problem into a
monolayer problem with spin \cite{Mac}.  Such a mapping allows one
to borrow for  double layer systems, the rich body of insights and
results available from single layer systems with real spin. 

Thus one may expect a fully symmetric (polarised) pseudospin
state to be energetically preferred because of a combination of
Coulomb repulsion and the Pauli principle which forces an
associated antisymmetric spatial wavefunction, just as in
itenerant ferromagnetism. Further, the relevance of Skyrmions to
systems with real spin, predicted by theoretical considerations
\cite{Sondhi}, \cite{Fertig} and supported by   experimental evidence 
\cite{Barrett}, has in turn prompted studies of similar topological
excitations in spinless double layer systems, but now involving
 pseudospin  (See Girvin and MacDonald
\cite{GirvMac}, Moon ${\it et al}$ \cite{Moon} and references given therein).

Because of interplane-intraplane anisotropy in Coulomb repulsion
between electrons located in the two layers, as well the
capacitance energy of maintaining unequal charge density in the
two layers, the effective Action governing pseudospin enjoys
only U(1) symmetry of rotations about the z-axis (the direction
perpendicular to the x-y plane of the layers).  Finiteness of
the capacitance energy between the two layers requires that
asymptotically the pseudospin must lie on the easy (x-y) plane.
The basic topological excitations in that case are the so-called
merons which are vortices in pseudospin with a winding number of
one-half (with respect to the second Homotopy group $\Pi_{2}$.
These are similar to vortices in the X-Y model, but non singular
at the origin since the pseudospin is not restricted to lie on
the x-y plane. But like the former they do have an energy that
grows logarithmically with size.  One can also have meron
anti-meron bound pairs whose energy is finite.  Such a pair is
topologically equivalent to Skyrmions and carries unit winding
number. (For an introduction to such topological excitations,
their winding numbers, etc. see reference \cite{Raj}.)

The possibilty of topological excitations like merons and
bimerons in double layer systems has generated much interest, in
part because of the excitement surrounding the Skyrmion
excitations in systems with real spin, and in part because of
the additional possibility here of Kosterlitz-Thoulles type
\cite{KT} phase transitions caused by the break-up of bound bimerons into
separated meron pairs \cite{GirvMac},\cite{Moon}.  Bimeron
solutions have already been extensively studied in a body of
papers by Girvin, MacDonald and co-workers \cite{Brey}
\cite{Yang} and \cite{Moon} . These calculations are based on 
optimising microscopic wavefunctions with respect to the
microscopic interaction Hamiltonian.  

We will also calculate bimeron solutions and their energies
here, but by using an alternate method. An Effective Action for
slowly varying pseudospin textures has already been obtained by
Moon et al \cite{Moon}. If one extremises that Action one will
get differential equations which the unit-vector valued field of
psuedo spin configurations ${\vec m}(\vec r )$ should obey in
the classical limit. 

In this paper we solve these coupled non-linear differential equations,
through a combination of analytically motivated ansatze followed
by numerical calculations. We obtain bimerons as approximate
time-independent solutions with appropriate topologically
non-trivial boundary conditions, for a range
of separations between the meron and its partner the anti-meron
and also for a set of different inter-layer distances.  The
dependence of the bimeron texture on these variables is
discussed.  They turn out to be reasonably similar to
 what one would expect on
general grounds.  We also obtain the energy of this bimeron as a
function of the separation between the meron centers. We 
include in this energy contributions coming from the
 pseudospin stiffness, its anisotropy, the capacitance energy and the
 Coulomb energy.  By minimising this energy with respect to
the meron separation, we are also able to give an independent
value for the optimal meron separation in a bimeron.  We compare
these results with earlier work, including our own.

Apart from this, our work also enables us to independently check
the validity of a physical picture often used \cite{Yang} in
estimating bimeron energies, namely, that they can be viewed as
a pair of rigid objects carrying electric charge of ${1 \over
2}$ and a logarithmically growing energy. A work somewhat
similar in spirit to ours, but  in the context of Skyrmions
of real spin systems was done by Abolfath ${\it et. al.}$ who
compared results obtained from solving a non-linear differential equation
with those obtained from microscopic calculations \cite {Abolfath}.
For yet another way of approaching meron solutions, starting from
a Chern- Simons field theory see the work of Ichinose and Sekiguchi
\cite{Ichinose}.

In an earlier paper \cite{Ghosh} we had done a similar study of
single meron solutions. But the present work is much more
complicated at the computational level. Single meron solutions are
circularly symmetric, with the spin component on the plane
pointing along the coordinate direction. Thus the only unknown,
namely, the spin-z component obeys an ordinary (though
non-linear) differential equation in the radius variable $r$.
Further, the boundary conditions relevant to a single meron can
be imposed, conveniently, at the end points $r=0$ and $r= \
\infty$. By contrast the boundary conditions characterising a bimeron
are $m_z = \pm 1$ at two finite points on the plane where the two merons
have their centers. The spin direction is also not simply related to
the coordinate direction, so that there are two independent
fields, say, $m_z$ and $tan^{-1}\bigg( {m_x \over m_y} \bigg)$,
(since the  ${\vec m}$ is constrained to be a unit vector)
which obey coupled partial differential equations on the plane.
We found it quite challenging to analyse these coupled equations 
analytically as far as posible, and use that information to 
employ an appropriate ansatz and coordinate system to numerically
 solve the equtions (using a desk-top computer).

Finally, we should reiterate that our work here clearly 
relies heavily on the advances already
made by the Indiana group \cite{Moon}, \cite{Yang}, \cite{Brey}
 and is to be viewed as something which will hopefully augment 
their findings.
 
\section{The Spin Texture  Equations .}

The differential equations obeyed by spin textures is obtained
by extremising an effective action which has already been
derived by Moon {\it et al} \cite{Moon} starting from the basic
microscopic physics. See also Ezawa \cite{z}. These results were
summarised in our earlier paper \cite{Ghosh}. Briefly , the
pseudospin texture of a state is described by a classical {\it
unit} vector ${\vec m}(\vec r )$ which gives the local direction
of the pseudospin. Here ${\vec r} $ is the coordinate on the x-y
plane carrying the layers, while the magnetic field B is along
the z-direction. The fully polarised "ferromagnetic" ground
state  corresponds to ${\vec m}$ pointing
everywhere in the same direction, say, along the x-axis.
 Using this as the reference state, any other state
with some arbitrary texture ${\vec m}(\vec r )$ is given by
performing a local pseudospin rotation on this uniform ground
state . The leading low-wavelength terms in the effective
Action for time independent configurations ${\vec m}(\vec r )$,
 as obtained by Moon {\it et al} \cite{Moon} is

\beq
I ({\vec m})=\int d^{2}r  \ \bigg[\frac{1}{2} \rho_{A} \big(\nabla
m_{z})^{2} + \frac{1}{2} \rho_{E} \big((\nabla m_{x})^{2} + 
(\nabla m_{y})^{2}\big) + 
  \beta  \ m_{z}^{2} \bigg] \ + \ C_{1}[m]  \ \ + \ C_{2}[m]  \ 
\label {Eff} \eeq 
where
\beq 
C_{1}[{\bf m}]  \ = \ \frac{1}{2}\int d{\vec r}d{\vec r'}V({\vec r}-{\vec r'})
q({\vec r})q({\vec r'})
\eeq

and 
\beq C_{2}[{\bf m}] \ \equiv {e^{2}d^{2} \over 32\pi^{2}\epsilon}
\int d^{2}r\int
d^{2}r'({m_{z}({\bf r})\nabla^{2}m_{z}({\bf r'}) \over |({\bf
r}-{\bf r'}|})
\eeq
 The constants $\rho_A$ and 
$\rho_E$ are pseudospin stiffness parameters  whose physical origin is the
exculsion principle (Hund's rule) mentioned earlier. They are given by
\beqarr \rho_A \ &=& \ \big( {\nu \over 32 \pi^2}\big) \int_{0}^{\infty} dk 
 k^3 \  V^A_k \ exp({-k^2 \over 2})    \nonumber \\
        \rho_E \ &=& \ \big( {\nu \over 32 \pi^2}\big) \int_{0}^{\infty} dk 
 k^3 \  V^E_k \ exp({-k^2 \over 2})   \label{rho}  \eeqarr
where $V^A_k \ = \ 2\pi e^2 /(\epsilon k)$ and  $V^E_k \ = \ 
(exp (-kd) 2\pi e^2) /(\epsilon k)  $
 are  the Fourier transforms of the Coulomb interactions between electrons
  in the same and different layers respectively.
  All distances (and inverse wave vectors) are in units of the
  magnetic length {\it l}. 
The $ \beta m_{z}^{2}$ term represents the so-called capacitance or
 charging energy needed to maintain unequal amounts of charge density
 in the two layers. Recall that the z-component of pseudospin represents
 the difference between the densities in the two layers. The constant
  $\beta $ is given by 
 \beq  \beta \ = \ \big( {\nu \over 8 \pi^2}\big) \int_{0}^{\infty} dk 
  \ k \ (V^{z}(0) - V^{z}(k)) \ exp({-k^2 \over 2}) \label{beta}   \eeq
where   $V^z_k = {1 \over2} (V^A_k - V^E_k)$. 
Finally, $q({\vec r})$ is the topological density associated with pseudospin
texture, which is also its charge density \cite{Sondhi}. It is given by
\beq 
q({\vec r})=-\frac{\bf\nu}{8\pi}\epsilon_{\nu \mu}{\bf m}({\vec r}).[
{ \partial_{\nu}}{\bf m}({\vec r}){\times}
{ \partial_{\mu}}{\bf m}({\vec r})]
\label{topo}\eeq
We will refer to the the non-local term $C_1$,  as the Coulomb term since
it has been identified as the Coulomb energy associated with topological 
structures in the pseudospin textures \cite{Moon}, \cite{Sondhi}. The
other non local term $C_2$ arises in the gradient expansion but is not
amenable to simple physical interpretation.

The field equations are obtained by extremising this Hamiltonian with respect 
to the independent field variables, which can be taken to be 
$m_z$ and
 $\alpha \equiv tan^{-1}\bigg( {m_x \over m_y} \bigg)$ . This $\alpha$ is
just the azimuthal angle of the projection of ${\vec m}$ on to the x-y plane.
 The non-local terms
$C_1$ ansd $C_2$ in the Action (\ref{Eff}) will render the field
equations into coupled integro-differential equations. While in the 
single meron
case we did solve such an integro differential equation \cite{Ghosh},
 for the more complicated
case of bimerons we will be content to solve the equations in the absence
 of the integral terms $C_1$ ansd $C_2$ .The contributions of these
terms can however be included in the total energy, but by using  solutions of 
the local equations. In mild justification of this strategy, we will
 find later that the Coulomb energy $C_2$ for instance is less than
half the energy from the local terms in eq. (\ref{Eff}).

  The coupled field equations for $m_z$ and
 $\alpha \equiv tan^{-1}\bigg( {m_x \over m_y} \bigg)$ resulting 
from eq. (\ref{Eff}) in the absence of $C_1$ and $C_2$ are
\beq
\rho_{A}\nabla^{2}m_{z} \ + \ \rho_{E} m_{z} \bigg( \frac{(\nabla
m_{z})^{2}}{(1-m_{z}^{2})^{2}}+\frac{m_{z} \nabla^{2}m_{z}}{1-m_{z}^{2}}+
\nabla^{2}\alpha \bigg) \  - \  2\beta m_{z} \ \\ =  \ 0 \label{mz} \eeq
and 
\beq {\vec \nabla} .\big[ (1-m_{z}^{2}){\vec \nabla} \alpha \big]=0 
\label{alpha} \eeq
\section {Bipolar coordinates}
To find bimeron solutions we have to numerically solve the 
coupled partial differential
equations (PDE) in (\ref{mz}) and (\ref{alpha}). 
The defining boundary condition of  a bimeron is $m_z \ = \ \pm 1$ at 
the points $(0, \pm a)$
Our strategy will be to use
the known exact solution of these equations in the Non-Linear Sigma Model
 (NLSM) limit, and solve the full equations
iteratively starting with the NLSM solution.
  The NLSM limit is realised when the layer separation $d$ goes
to zero in which case we see from their defining equations above that $ \rho_A
 \ = \ \rho_E $, i.e. the stiffness is isotropic and further that the
capacitance coefficient $\beta$ vanishes. Then , with $C_1$ and $C_2$ also
neglected, the action in (\ref{Eff}) is just that of the NLSM, all of whose
solutions are exactly known \cite{Raj}. They are conveniently described by
 the complex field   w(z) which represents the stereographic
projection of the unit sphere of textures ${\vec m}$. It is defined by
 \beq w(z) \equiv {m_x + im_y \over (1 - m_z)} \eeq
 where z = x+iy. 
Our texture variables $m_z$ and $\alpha$
are related to w(z) by
\beqarr m_z \ &=& \ {|w|^2 - 1 \over |w|^2 + 1}  \nonumber \\
and  \ \ \ \ \alpha \ &=& \ arg  \ (w) \label{mzw} \eeqarr
Any analytic function w(z) will be a solution of the NLSM.
 In particular the function
\beq w(z) \ = \ {z - a \over z + a} \label{NLSM} \eeq
represents the bimeron, with the points (0,-a) and (0,a) representing
the centers of the two merons, where the solution gives $m_z = \pm 1$
respectively. It may be checked that (\ref{NLSM}) satisfies the coupled
equations (\ref{mz}) and (\ref{alpha}) in the isotropic limit.

When the interlayer separation d is not zero, we have to cope with 
the coupled field equations (\ref{mz}) and (\ref{alpha}) with  both 
the anisotropic  stiffness and capacitance terms present. Some analysis
of this system was done long ago by  Ogilvie and Guralnik \cite{Ogil}
who studied the  NLSM  with the mass (capacitance) term included but
no anisotropy. ( An ansatz suggested
 in ref (\cite{Ogil}) does not work ,as we will show below.)
Meanwhile Watanabe and Otsu \cite{Wata} studied the anisotropic NLSM but
without the mass term. Both made considerable progress  analytically,
but neither offered  exact or numerical solutions. Here we
will try to solve (\ref{mz}) and (\ref{alpha}) numerically after
including both the capacitance and anisotropic terms . 

To do so , it will be convenient to use a bipolar coordinate system to
describe the x-y plane, as
 might be expected when we have to impose boudary conditions at two
finite points (0,-a) and (0,a). These coordinates, $\eta$ and $\phi$,
 are defined by
\beqarr  \ \ \ \eta \equiv \  \ ln  \ |{z-a \over z+a}| \nonumber \\
and \ \ \   \phi \equiv arg  \ \bigg( {z-a \over z+a } \bigg)  
\label{etaphi} \eeqarr 
This coordinate set has many advantages \cite{Margenau}. 
The points (0,-a) and (0,a) at which we have to impose
boundary conditions are now mapped into $\eta \rightarrow \pm
\infty $. The full x-y plane is mapped in $(\eta,\phi)$ coordinates to an 
infinite strip with $\eta = [-\infty, +\infty]$ and $\phi =
[-\pi, \pi]$. Finally, it is clear upon comparing eq(\ref{etaphi})
to eq (\ref{NLSM}) that this set of coordinates is
closely related to the exact NLSM bimeron solution. Clearly the 
 the exact NLSM solution (\ref{NLSM}) 
corresponds to the simple expressions
\beqarr m_z \ = \ tanh \eta  \nonumber \\
and \ \ \ \ \alpha \ = \  \phi \eeqarr 

Away from the NLSM limit, since this is an 
orthogonal coordinate system with simple expressions for the 
gradient, divergence and Laplacian, 
the equations (\ref{mz}) and (\ref{alpha}) become
\beqarr
\ \bigg[(\frac{ \rho_{A}-\rho_{E}}{\rho_{E}}) + \frac{1}{1-m_{z}^{2} }\bigg]
 (\partial_{\eta}^{2}m_{z} +\partial_{\phi}^{2}m_{z}) +\frac
{m_{z}({\partial_{\eta}m_{z} +\partial_{\phi}m_{z}})^{2}}{({1-m_{z}^{2}})^{2}}
+m_{z}(({\partial_{\eta}\alpha +\partial_{\phi}\alpha})^{2}
 \nonumber \\
-  \frac{2\beta}{\rho_{E}}  \ \ Q^{2} \ (\eta, \phi)  = 0  \label{mz1}\eeqarr

\beq (1-m_{z}^{2})(\partial_{\eta}^{2}\alpha +\partial_{\phi}^{2}\alpha)
-2m_{z}({\partial_{\eta}m_{z} \partial_{\eta}\alpha +\partial_{\phi}m_{z}
\partial_{\phi}\alpha}). =0 \label{alpha1} \eeq

where
 \beq Q^{2} \ (\eta, \phi) \ =  \frac{a^{2}}{({\cosh{\eta}-\cos{\phi}})^{2}}
 \eeq
is the Jacobian of this coordinate transformation.
Now let us analyse these equations as different terms are included in stages.

(a)  In the NLSM limit,   our exact solution has $\alpha = \phi$.
Then (\ref{alpha1}) forces $m_z$ to be a function of $\eta$ alone ,
$m_z = m_z \ (\eta)$. Upon inserting this into the other equation (\ref{mz1})
it becomes an {\it ordinary} non-linear differential equation . This
is the advantage of this choice of coordinates. The
solution can be verified to be $m_{z} = tanh(\eta)$.

(b) Next let us include anisotropy $( \rho_A \neq \rho_E)$ ,
 while still keeping the capacitance term zero $(\beta = 0)$.
Once again we can set $\alpha = \phi$, and consequently $m_z = \ 
m_z (\eta) $, which will obey again an ordinary differential
equation given by   
\beq
\ \bigg[(\frac{ \rho_{A}-\rho_{E}}{\rho_{E}}) + \frac{1}{1-m_{z}^{2} }\bigg]
 (\partial_{\eta}^{2} \ m_{z} ) +\frac
{m_{z}(\partial_{\eta}m_{z} )^{2}}{(1-m_{z}^{2})^{2}}
+ m_{z} \\  = 0  \label{mz2}\eeq

This has no analytic solution, but can be solved relatively
easily numerically, being just an ordinary differential equation in
the variable $\eta$. As boundary conditions we impose
$ m_z = o$ \ \  \  at $\eta = 0$ and $m_z = 1$ at $\eta = \infty$,
 (Note that the equation above
is symmetric under $\eta \rightarrow  \ - \eta$, so that we can choose the
solution to be antisymmetric, i.e. $ \  m_z (- \eta) = - m_z( \eta)$).
The resulting numerical solutions for different values of layer 
separation $d$ (on which the anisotropy depends), are shown in fig 1.
One can see that with increasing the layer separation, and hence
increasing anisotropy in the stiffness, the pseudospin component
 $m_{z}$ reaches its asymptotic value more slowly.

(c) Finally let us also include the capacitance term and consider
 the equations
(\ref{mz1} and \ref{alpha1}) in full. Now the ansatz $\alpha = \phi$ is
no longer sustainable, in contrast to what has been suggested in
ref (\cite{Ogil}). The substitution of the ansatz $\alpha = \phi $ in
 equation (\ref{alpha1}) would again force $\partial_{\phi} m_z = 0$  ,i.e.
$ m_{z} = m_{z}(\eta) $. But now this is in  contradiction with 
equation (\ref {mz1}) which has an explicit $\phi $ dependence
through the last (capacitance) term  
$\frac{2\beta}{\rho_{E}} \ \ Q^{2} \ (\eta, \phi)$. Therefore ,once
one includes the capacitance term in  equation(\ref{mz1}) 
both $\alpha$ and $m_{z}$ become functions of both $\eta$ and $\phi$.
One has unavoidably to 
solve the coupled non-linear PDE  for $m_z = m_{z}(\eta,\phi)$ and 
$\alpha= \alpha(\eta,\phi)$. 

We do this by employing what we believe is a 
 good ansatz for $\alpha$ which approximately satisfies \ref{alpha1}).
 We then  solve the other equation (\ref{mz1}) numerically after 
inserting that ansatz for $\alpha$.
Our ansatz is been motivated by the following arguments.
One can see from  equation(\ref{mz1}) that the troublesome $\phi$ dependent
term $Q^2$ is negligibly small in the large $\eta$ region
 $(Q \sim sech (\eta))$
and is most dominant in the small $\eta$ region.
Hence $\alpha$ will still approach  $\phi$ as $\eta
\rightarrow \infty$ but needs to be modified substantially 
 in the small $\eta$ region where however $m_{z} \ll 1$.
When $m_{z} \ll 1$ equation (\ref{alpha1}) can be approximated by
\beq \bigtriangledown^{2}\alpha=0 \label{laplace}\eeq
This is just Laplace's equation in two dimension whose solutions
are all harmonic functions. With this in mind we choose  our ansatz for
$\alpha$ as follows :
\beq
\alpha = \phi -B \kappa exp(-|\eta|)sin(\phi)
\label{alpha2}\eeq
where
\beq 
\kappa \equiv (\frac{2\beta}{\rho_{E}})^{1\over 2} \ \ a \ \eeq 

This  solves  Laplace's equation 
and satisfies all the required boundary conditions and asymptotic behaviour,
namely
\begin{eqnarray}
\alpha \rightarrow \phi \ \ \ \  &as \ \ \ \  \eta \rightarrow
 \pm \infty \nonumber \\
\alpha = 0  \ \ \  &when   \ \ \ \ \ \phi =0 \nonumber \\
\alpha = \pi  \ \ \ &when  \ \ \ \ \phi =\pi \nonumber \\
\alpha = \phi  \ \ \ &when  \ \  \ \kappa =0 \label{boundary} \end{eqnarray}.
Note that the ansatz has a cusp at $\eta = 0$. This need not cause
concern. Some such cusps can be
expected on physical grounds and are familiar in soliton physics. The
point is that  each meron feels some force due to the other (Coulomb
plus a logarithmic force) at arbitrary separation. We would expect them
to move because of this force, and cannot strictly
speaking expect a static bimeron solution to exist at arbitrary
separation.  But a cusp, like the
one in the above ansatz, amounts to a delta function in the second
derivative and can be interpreted as a external force just at $\eta = 0$ 
which can "hold the two merons together" at arbitrary separation. For
more discussion of this point see Rajaraman and Perring and Skyrme
\cite{RR} where this technique was used to get intersoliton forces betwen
one dimensional solitons. 

The constant B is chosen by minimising the energy.
Substituting this ansatz in  equation (\ref{mz1}) we then solved it
numerically subject to the boundary condition 
\begin{eqnarray}
m_{z} \ = 0  \ \ \ \ &at  \ \ \ \eta =0\nonumber \\
m_{z} \ = \pm 1  \ \ \ &when \ \ \ \eta = \pm \infty \label{kboundary}
 \end {eqnarray}.
It is sufficient to solve the equation in the first quadrant .i.e.
$(\eta [0,\infty] and \phi [0,\pi])$ . For the rest of the
quadrants solutions can be obtained by writing
\begin{eqnarray}
m_{z}(-\eta,\phi)=-m_{z}(\eta,\phi) = -m_{z}(\eta,-\phi)\nonumber \\
\alpha(-\eta,\phi)=\alpha(\eta,\phi)=-\alpha(\eta,-\phi) \nonumber \\
\end{eqnarray} 
which is consistent with the invariance of equations
(\ref{mz1})and(\ref{alpha1})  under the transformation $\eta
\rightarrow -\eta$ and $ \phi \rightarrow -\phi$.

\section{Numerical Procedure}

Before proceeding to solve this PDE (\ref{mz1}) we must take note of the
fact  that the last term of the equation 
(\ref{mz1}) is singular at the point 
$(\eta=0,\phi=0)$. This point corresponds to spatial infinity
on the parent x-y plane.
As one moves near this point the leading
singularity in the equation, coming from the $Q^2$ term,
 goes like $\frac{4\kappa^2}{(\eta^{2} +\phi^{2})^{2}}$
with other subleading singuarities of the form 
$\frac{1}{\sqrt(\eta^{2} +\phi^{2})}$. It can be seen that this leading
singularity can be offset by requiring that  $m_{z}$ behave as
$\bigg[exp-\bigg(\frac{2\kappa}{\sqrt{\eta^{2}+\phi^{2}}}\bigg)
\bigg] \ g (\eta,\phi)$, where this $g (\eta,\phi)$, 
is a more smooth function for which one solves numerically.
This corresponds,  in more familiar polar coordinates $(r,\theta)$ to
writing $m_z$ in the form $[exp-(\frac{\kappa r}{a})] $
 \ ${\tilde g} \ (r,\theta)$. That $m_z$ will suffer
 such an exponential fall-off
as $r \rightarrow \infty $ can also be inferred directly from the 
"mass term " $2\beta m_z$ in the original field equation (\ref{mz}).
Similarly one can also verify that the
cancellation of the subleading singular terms can be achievedd by
requiring that $g$ has to behave like
 $\sqrt(\eta^{2}+\phi^{2})$ as $\eta, \ \phi \ \rightarrow 0$.

Given this functional form of $m_{z}$ near the origin of the $ \eta, \phi$
plane, the boundary conditions (\ref{kboundary}), and  the 
ansatz (\ref{alpha2} )for $\alpha$  we solved equation
(\ref{mz1}) through an iterative procedure. We start with the solution
for $\kappa$=0 but with full anisotropy, which can be obtained relatively
easily from  the ordinary differential equation (\ref{mz2}). 
We then use this solution as input to obtain the
solution for $\kappa $ equal to a small number $\epsilon$ 
through the Newton-Raphson method
\cite{numer}. The solution for $\epsilon$ is then used as input 
to obtain the solution  for $2\epsilon$ and so on. This procedure
 is repeated until one reaches the desired value of $\kappa$. The
 advantage of this
procedure is that one can make $\epsilon$ arbritrarily small
to make the Newton-Raphson method converge. In this way we  obtained
 solutions  for different values of the ansatz parameter B
 for each value of $\kappa$.

\section{RESULTS and DISCUSSION}
Our solutions of equations(\ref{mz1}) and along with the ansatz
(\ref{alpha2}) give us the value of the pseudospin vector ${\vec m}$ as a
function of $\eta$ and $\phi$, or equivalently, the value of the vector-field 
${\vec m}$ on a lattice of points on the parent x-y plane. We repeated
 this calculation for a set of values
of the paramenter B in the ansatz (\ref{alpha2}).
 We found that as one varies B starting from 0, the energy does not vary much
as B goes 0 to 0.1, but then it increases sharply after
B=0.1. This behaviour is seen to be common to all $\kappa$ and all $a$. 
Hence we take B to equal 0.1. and solve the PDE for a variety of 
values of layer separation {\it d}, and bimeron separation {\it a}
({\it a} is actually half of the meron-antimeron separation) . Together
all these solutions represnt a large body of calculated data. But it is
neither feasible nor very interesting to try to display it all in this
paper. Instead we will try to bring out salient features of our solutions
through examples.

Recall from (\ref{mz2})that in the absence of the capacitance term $m_z$
 had no $\phi$ -
dependence. To give some feel for how the $m_z$ varies with $\phi$ in the
presence of the capacitance term, we plot in fig. 2 the solution
$m_{z}(\eta)$ of equation (\ref{mz1}) for a set of values for $\phi$. This
solution corresponds to $ d= 0.7$ and $a=3.158$. The sequence of curves shown
correspond to $\phi$ equal to 0, 0.2$\pi$, 0.47$\pi$, and 0.94$\pi$
respectively with the outermost one belonging to $\phi$ equal to 0. As we
have discussed earlier, as $\eta$ and $\phi$ tend to zero, the solution
should damp exponentially as
$exp(-\frac{\kappa}{\sqrt{\eta^{2}+\phi^{2}}})$. Correspondingly we see in
fig.2 that the low $\phi$ curves rise very slowly as $\eta$ increases away
from zero. We also give for comparison, in the form of the dotted curve,
the function tanh ($\eta$) which is the solution in the NLSM limit.  The
comparison shows that the restructuring of the pseudospin texture due to
the capacitance and term and anisotropy is considerable.

As an alternate representation of our results, we show in  fig. 3
the projection of ${\vec m}$ on the x-y plane, for the example of 
$d$ equal to
0.7 and $\kappa$ equal to 4.4. (All lengths throughout this article are
in units of the magnetic length ${\it l}$). The length of each
arrow gives the magnitude of its easy-plane projection 
$\sqrt{m_{x}^{2}+m_{y}^2}$ 
and its direction gives the azimuthal angle 
of the projected vector , namely, $\alpha =  \ 
tan^{-1}\bigg(\frac{m_{y}}{m_{x}}\bigg)$. 
The plot clearly shows that our " bimeron" solution is indeed
 a meron-antimeron pair. Note that, as desired, 
 $\vec m$ lies along the x-axis asymptotically. This picture
closely resembles the general structure obtained by Brey {\it et.al.}
 \cite{Brey}. The data corresponding to all other values of $d$ and $a$ 
we studied have a similar behavior. 

In fig.4 we plot the topological charge density given in eq.(\ref{topo})
as a function of $\eta$ and $\phi$ in the presence of all the local terms
in the field equations, including anisotropic ones. In viewing 
this figure it may be
helpful to remember that large $|\eta|$ corresponds to the meron centers
while  $\eta = 0, \phi=0$ corresponds to spatial infinity.  $ \phi=
\pi $ corresponds to the line joining the two merons.As topological
charge density is symmetric when either of the co-ordinate variable
changes sign we  show the contours only in the first quadrant where both
$\eta$ and $\phi$ are positive.

Next let us turn to the energetics of these bimeron solutions.  In fig.5. 
we show how the "local" energy i.e. the  contribution  from the local
terms in the energy functional (all terms in eq(\ref{Eff}) except
for $C_1$ and $C_2$) varies when one changes the separation $2a$
between the meron and antimeron centres. The appearance of a minimum
is quite conspiquous and generic to all the layer separations for
which the energy is calculated. The example in fig. 5 corresponds 
to a layer separation of 0.7.

In fig.6 we plot the Coulomb energy $C_{1}$ evaluated using our solution
of the equation (\ref{mz1}), as a function of the bimeron separation. The
continuous curve is the best fit to our calculated points . Sometimes in
the literature, a phenomenological estimate of bimeron energetics is made
assuming that it can be viewed as a bound pair of two merons, each
symmetrical, undistorted by the other and carrying a charge of
$\frac{e}{2}$ . Such a pair would have a Coulomb energy of $\frac{1}{8a}$
(in units of $\frac{e^2}{\epsilon {\it l}}$ that we are using). To see how
good an approximation this simple picture is, we give in the same fig.6,
in the form of a broken line the plot of this function $\frac{1}{8a}$ 
. We see that the value of the Coulomb energy we get from the actual bimeron 
solution is much larger than what the simple two-charge picture would
give. This is presumably because each meron is considerably squashed
(polarised) by the close proximity of the other. In our earlier work on
single merons \cite{Ghosh}, we had found that at the layer separation
($d=0.7$) used in fig.6, the core-radius of individual merons is about 2,
which is of the same order as the meron-separation in fig 6. In fact we
can see that the gap between the two curves in fig.6 is higher for smaller
$a$ where the individual merons are squeezed together more. Of course our
results, while indicative , may not be quantitatively unambiguous.  For
instance, recall that our solution was obtained using only the local terms
in the differential equation and the Coulomb energy was calculated by
substituting this solution into the integral $C_1$.  The non-local Coulomb
term's influence ${\underline on}$ the solution has no been included.

In fig.7 we plot the variation of three terms in the energy fuctional
namely the contribution from the local terms (capacitance+gradient energy)
,$C_{1}$ and $C_{2}$ as a function of the bimeron separation.The data
presented here corresponds to layer separation $d$ equal to 0.7$\it l$
but this behaviour is representative of almost all the layer separations 
( 0.5, 0.6, 0.7 and 0.8 for which we have found solutions. 
The  trend of all three contributions is the same
for the other layer separations also with only slight changes in the slope of 
the curves. 

Our calculations were done for different bimeron separations
$a$, for each layer separation $d$. In reality, the exact
solution should exist only for some optimal bimeron separation
$a$ for each value of $d$.  One can ask if our calculations
would reveal this by minimising the total energy at some
particular $a$. To see this, we have shown in Fig. 8 the total
energy at d=0.7 (i.e. the sum of all three contributions plotted
in the fig.7) as a function  of bimeron separation $a$.  As we
can see from fig.7, the total energy keeps decreasing with $a$,
all the way to  about $ a = 3.2$, which is the highest value
upto which we could calculate , given limitations of our
computing facilties. However, the decrease is clearly levelling
off and is indicative that a minimum may exist at around a=4 or
5. What we have done, in drawing fig.8,  is to obtain a
best-fit-curve of the data points upto a=3.2 and extrapolate
that curve upto a=4.5. For what it is worth such extrapolation indicates a
minimum at about a=4.  This corresponds to a meron-antimeron separation of
 of about 8, larger than what Yang and
MacDonald found by entirely different methods (see their fig. 2)
\cite{Yang}. Their value of the meron separation for $d=0.7$ is
about 4.5. We attribute this discrepency to the fact , noted
already in our discussion of fig.6, that the Coulomb energy in our
explicit calculation of the bimeron solution is higher than
the undistorted meron pair  estimate used in ref(\cite{Yang}).
The actual larger Coulomb repulsion is, we believe responsible for the 
larger optimal meron separation that we get.

We saw that the Coulomb interaction energy between the two merons as given
by the term $C_1$ in the present calculation differs quite a bit from the 
simple picture of the bimeron as a pair of undistorted merons of charge $
\frac{e}{2}$ each. One can ask if there is a similar discrepency in the 
non-Coulombic energy as well. This is the subject of Table 1. 
In the picture of a bimeron as a pair of merons  \cite{Moon} ,
 \cite{Yang},  \cite{Ghosh}, it will have energy equal to
\beq E_{prev}  \equiv \ 2E_{mc} + \ 2\pi \rho_{E}  \ \ ln \bigg(
\frac{2a}{R_{mc}}\bigg)
\eeq
 where $E_{mc}$ and $R_{mc}$ are respectively the core energy 
and radius of a single merons, which have a   logarithmic interaction 
with each other because of the logarithmic divergence of the self energy
of single merons. (As stated already we are leaving out their
Coulomb interaction in the comparison being done in this table.)
This $E_{prev}$ has been calculated in our previous work
\cite{Ghosh}. It can be compared with the local part of the
energy in the present calculation. Such a comparison is given in
Table 1 for different values of $d$, using the optimal value of
the meron separation $a$ which minimises $E_{local}$. We see
that the comparison is not bad considering the completely
different ways of estimating this energy in this paper and in
earlier literature.

In conclusion, our  solution for  the  bimeron obtained 
by directly solving the coupled partial differential equations that  
the bimeron texture obeys provides an alternate way of
obtaining the profiles and energies of these objects. As far as
the local part of the energy is concerned, the results are in
broad agremment with microscopic derivations earlier.  But the
Coulomb energy we obtain is higher by a factor of about 2 from earlier 
simple estimates because in actuality, the two merons in close
proximity will not behave like undistorted symmetrical merons.

\section{Acknowledgements} We are indebted to Awadesh Prasad for 
his unstinting help on many fronts. SG would also like to thank Sujit
Biswas and Anamika Sarkar for helpful discussions on the numerical work. 
SG acknowledges the support of a CSIR Grant no.
9/263(225)/94-EMR-I.dt.2.9.1994.

\begin{figure}
\label{fig1}
\caption{The solution $m_{z}(\eta)$ of equation (\ref{mz2}).
The three continuous curves correspond, as you go outwards, to three different
values of layer separation $d$ equal  to 0.5, 0.6 and 0.7 respectively
in the unit of magnetic length ${\it l}$.
The dotted curve corresponds to the exact solution of NLSM i.e.
$m_z =tanh(\eta)$.}
\end{figure}
\begin{figure}
\label{fig2}
\caption{The solution $m_{z}(\eta)$ of equation (\ref{mz1})
for a set of values for $\phi$.The curves correspond, as you go inwards, to
$\phi = 0 , 0.2\pi, 0.47\pi, 0.94\pi$ respectively with 
the outermost one corresponds to $\phi$ equal to 0.The layer separation
 $d$ is equal to 0.7{\it l} and bimeron separation $a$ is equal to 
3.158{\it l}. 
The dotted curve at the top again corresponds to
$m_z =tanh(\eta)$.}
\end{figure}
\begin{figure}
\label{fig3}
\caption{This figure gives  the magnitude and direction 
of x-y projection of $\bf m$  at different points on the plane.
The layer separation and the  bimeron separation are same as in
fig. 2.}
\end{figure}
\begin{figure}
\label{fig4}
\caption{A contour plot of the  topological charge density of
the bimeron when both the capacitace term and the anisotropy
term is incorporated.This particular plot corresponds to a
layer separation $d$equal to  0.7 and bimeron separation $a$ 
equal to 3.158 both in the unit of magnetic length {\it l}.  
The number against each contour(shown by broken curves) 
denotes the corresponding charge density.}
\end{figure}
\begin{figure}
\label{fig5}
\caption{This figures gives the plot of the energy($E_{local}$) 
coming from the 
local terms in the action as a function of bimeron separation $a$
in the unit of magnetic length ${\it l}$. The unit of energy is 
$\frac{e^{2}}{\epsilon {\it l}}$. The points correspond to the actually
computed values of the energy while the  continuous curve is the best
fitted curve to it. The form of the best-fit curve is $E = A + B
(a-C)^{2}$ where A and B and C are found out to be 
.223,.008 and 2.76 respectively.
This data corresponds to a layer separation $d$ equal to $0.7{\it l}$}
\end{figure}
\begin{figure}
\label{fig6}
\caption{This figure gives the plot of the coulomb energy as a function of
bimeron separation $a$ in  units of magnetic length ${\it l}$.
The unit of energy is $\frac{e^2}{\epsilon\it l}$. The upper
curve is our computed value of the coulomb energy integral $C_1$
using the solution of equation (\ref{mz1})(points). The contious
line is the best curve to these points. The form of the best fitted
curve is $E = \frac{A}{a^{B}}$ where A and B are found out to be 0.847
and 0.821 respectively. 
The dotted curve 
at the bottom corresponds to the  Coulomb energy
that   the bimeron would have, if viewed as a bound pair of two 
point charges of $\frac{e}{2}$ each, separated by a distance 
$2a$ . This data corresponds to a layer separation
$d$ equal to 0.7}
\end{figure}
\begin{figure}
\label{fig7}
\caption{this figure gives a relative estimate of the contribution
of the three type of terms  in the action, namely, the local terms,
$C_{1}$ and $C_{2}$,  as a function of bimeron separation $a$. The units
are as specified in the earlier figures.This data also corresponds 
to a layer separation $0.7{\it l}$} 
\end{figure}
\begin{figure}
\label{fig8}
\caption{A plot of the total energy $E(total)$
as a function of bimeron separation $a$, for a 
 layer separation of 0.7. 
 This curve was obtained by extrapolating  the curve fitted to the
calculated values going upto $a = 3.2$ . }
\end{figure}
\newpage

\noindent Table 1:  The optimal bimeron  separation
 ($a$), the bimeron local energy( $E_{local}$ ) and meron 
 pair energy ($E_{prev}$) from our previous work \cite{Ghosh} as
  a function of the layer separation $d$.
The unit of energy is $\frac{e^{2}}{\epsilon l}$ and the unit of length
is ${\it l}$

\begin{center}
\begin{tabular}{|c|c|c|c|}
\hline
d &  $a$ & $E_{local}$ & $E_{prev}$ \\  
\hline
0.5 & 3.30 & .270 & .217 \\
\hline
0.6 & 3.16 & .248 & .226 \\
\hline
0.7 &2.72 & .223 & .224 \\
\hline
0.8 & 2.39 & .201 & .214 \\
\hline
\end{tabular}\\
\end{center}

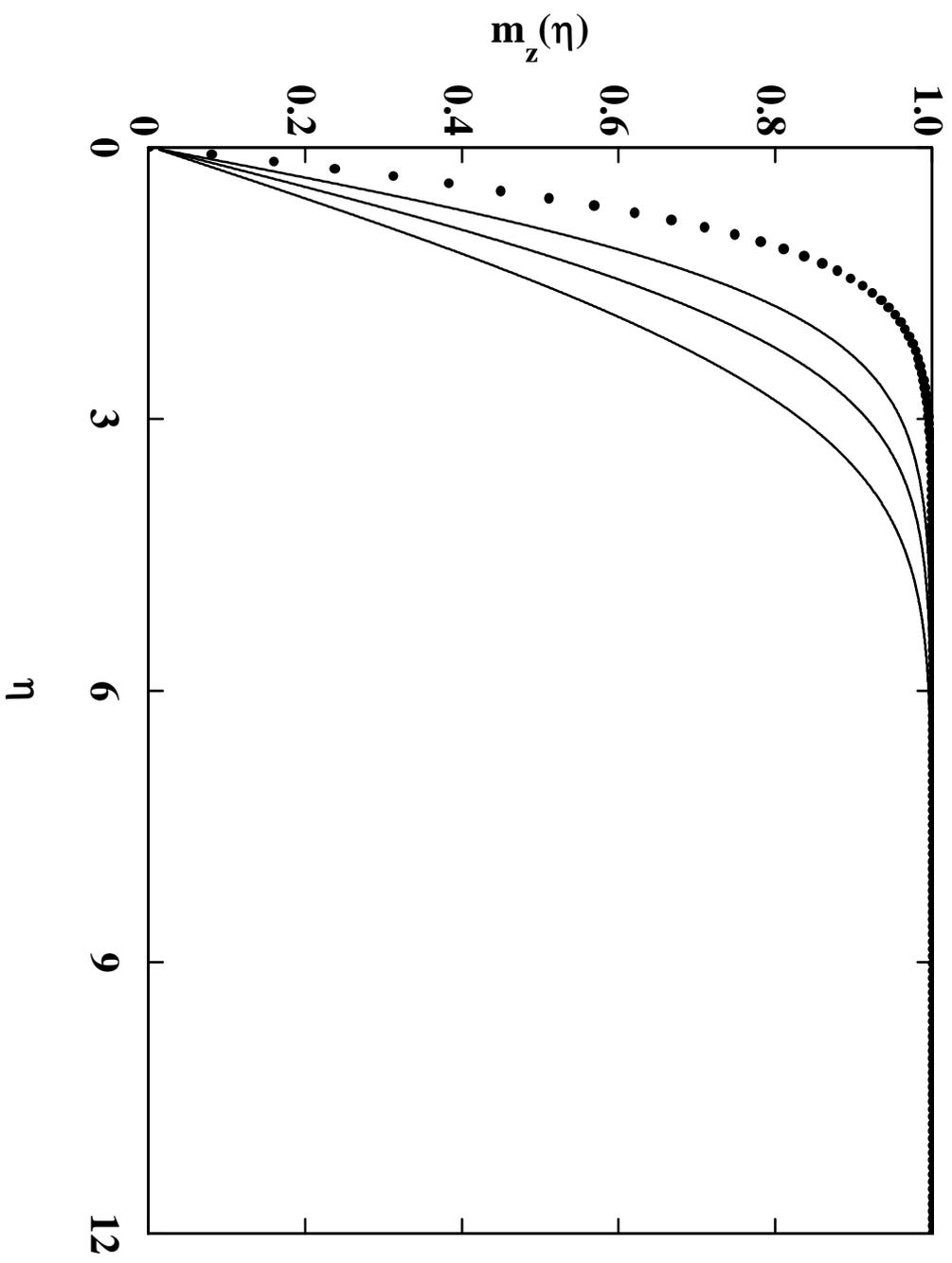
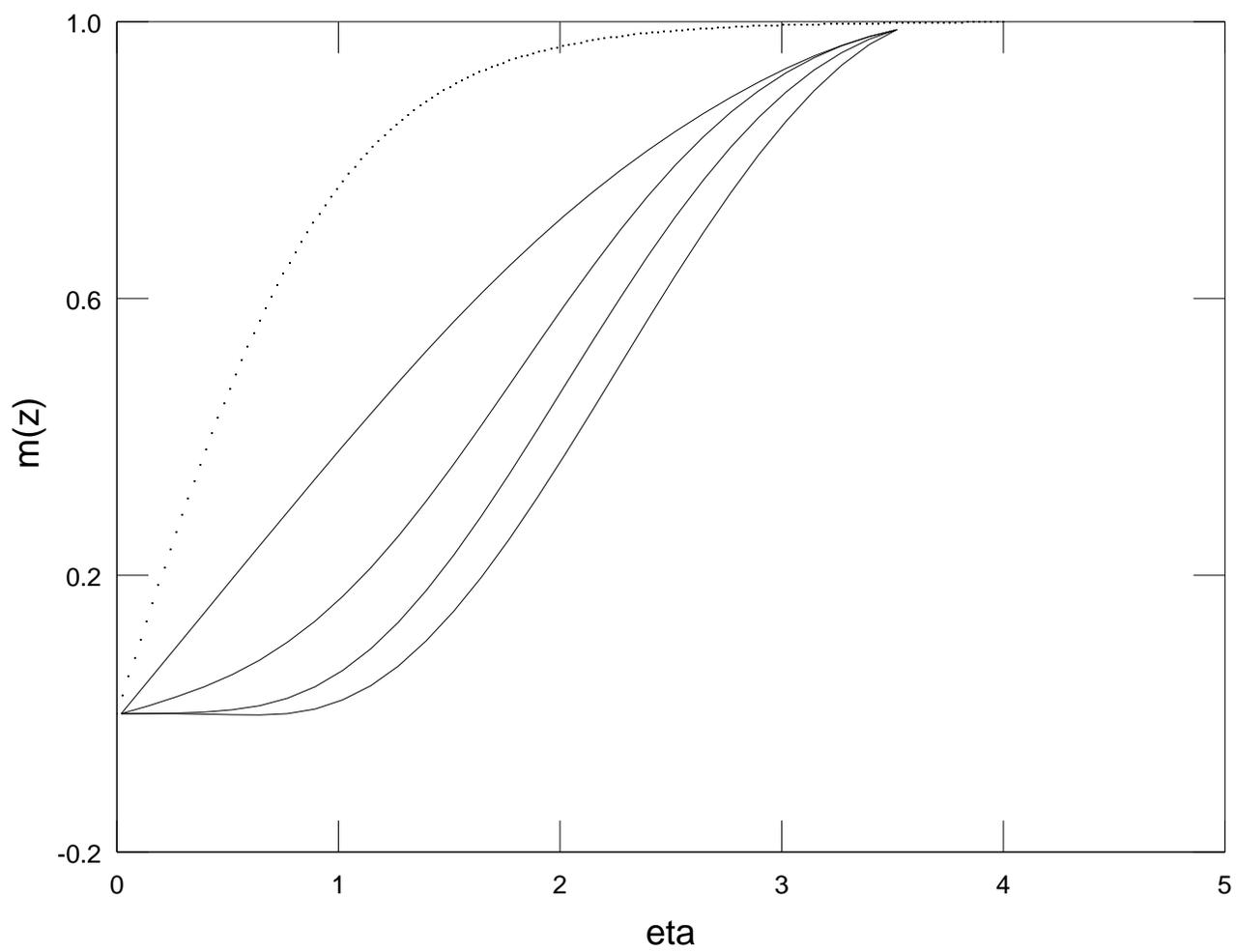
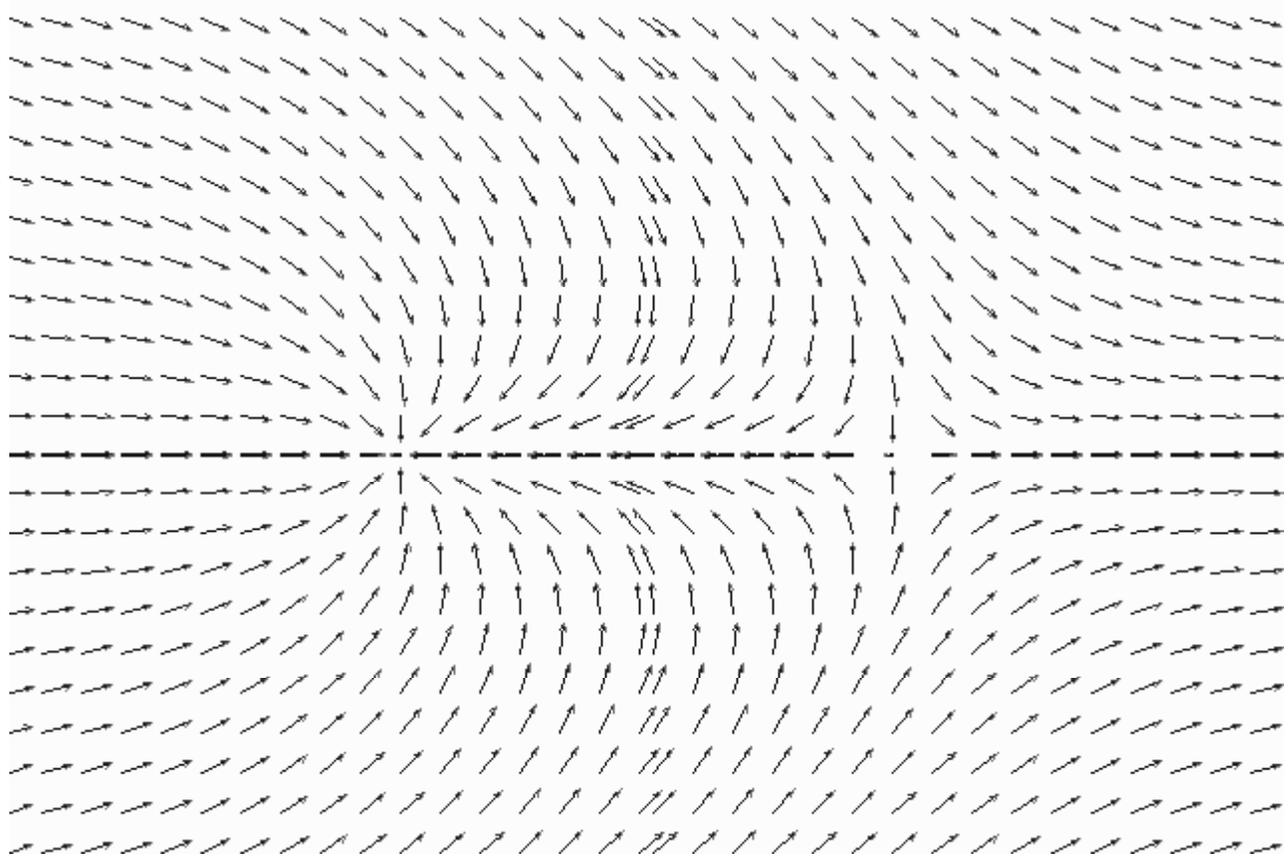
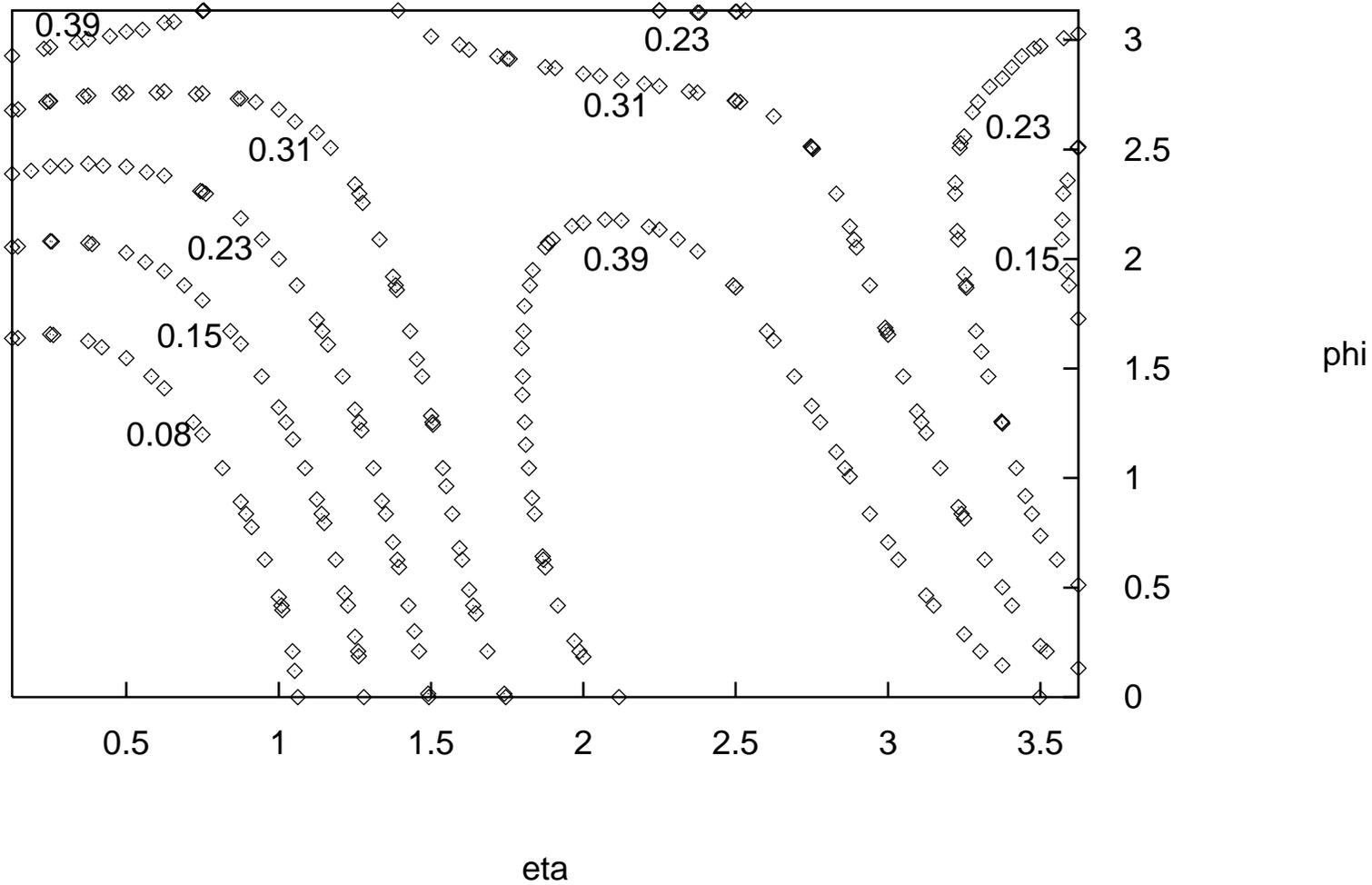
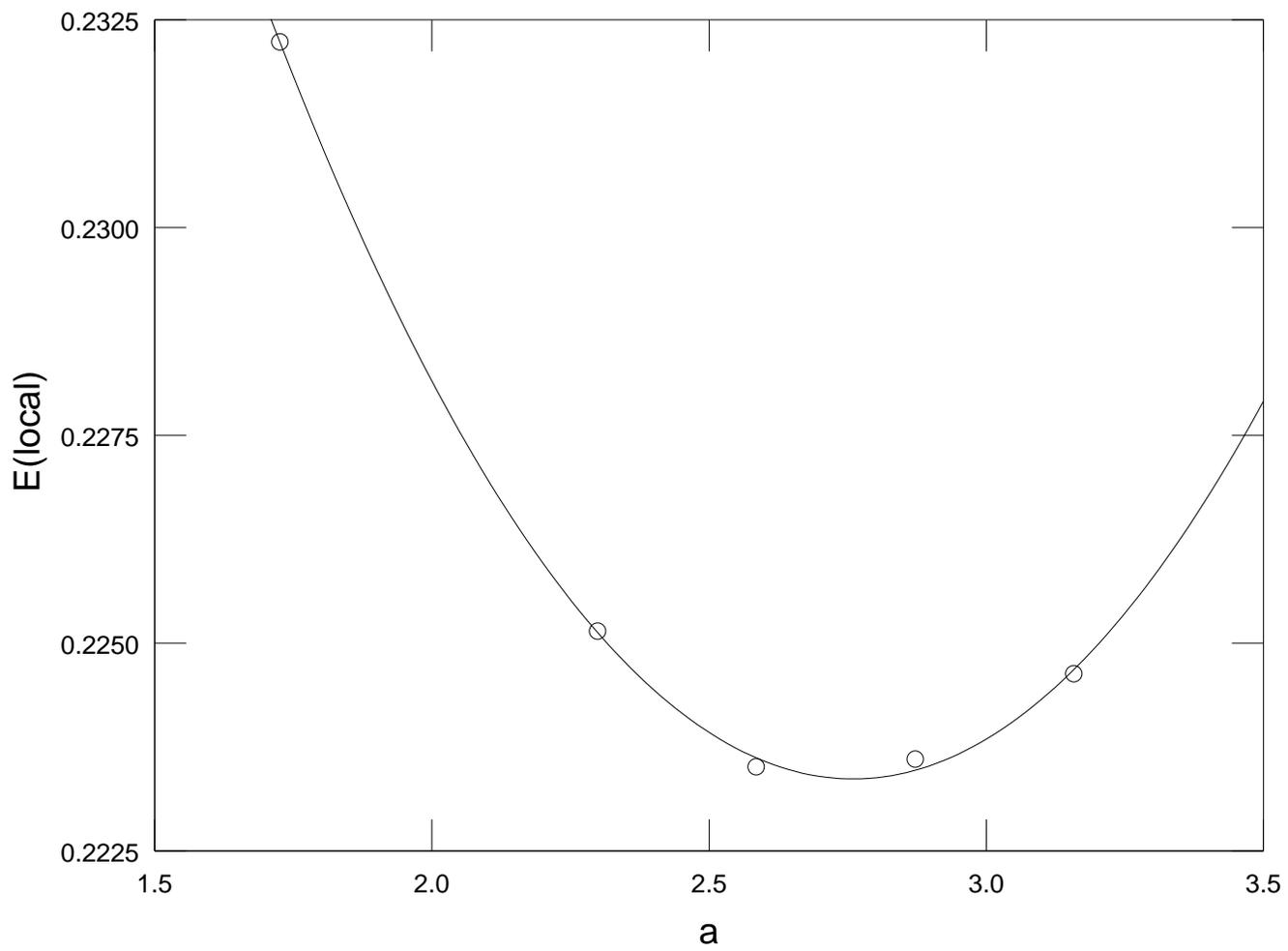
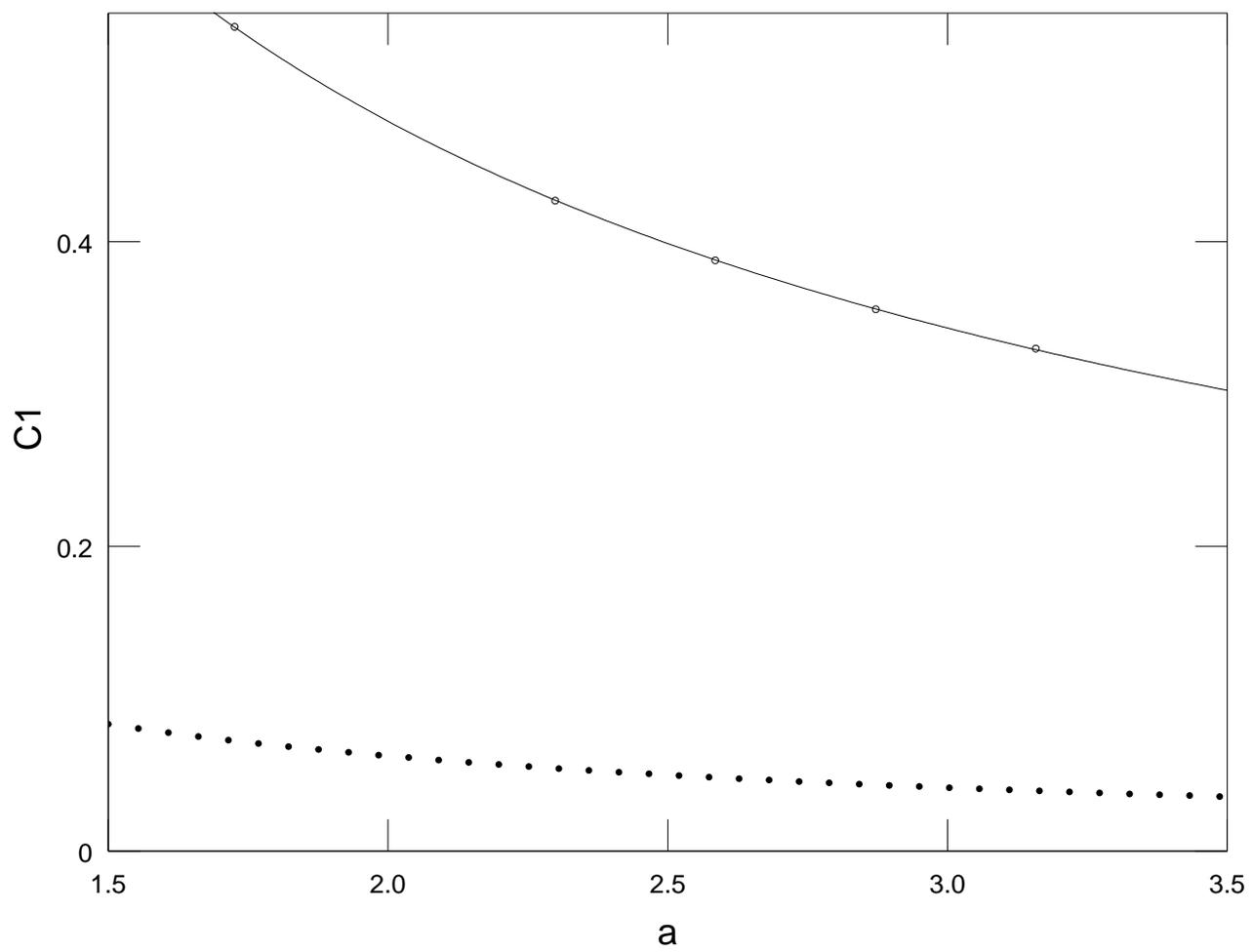
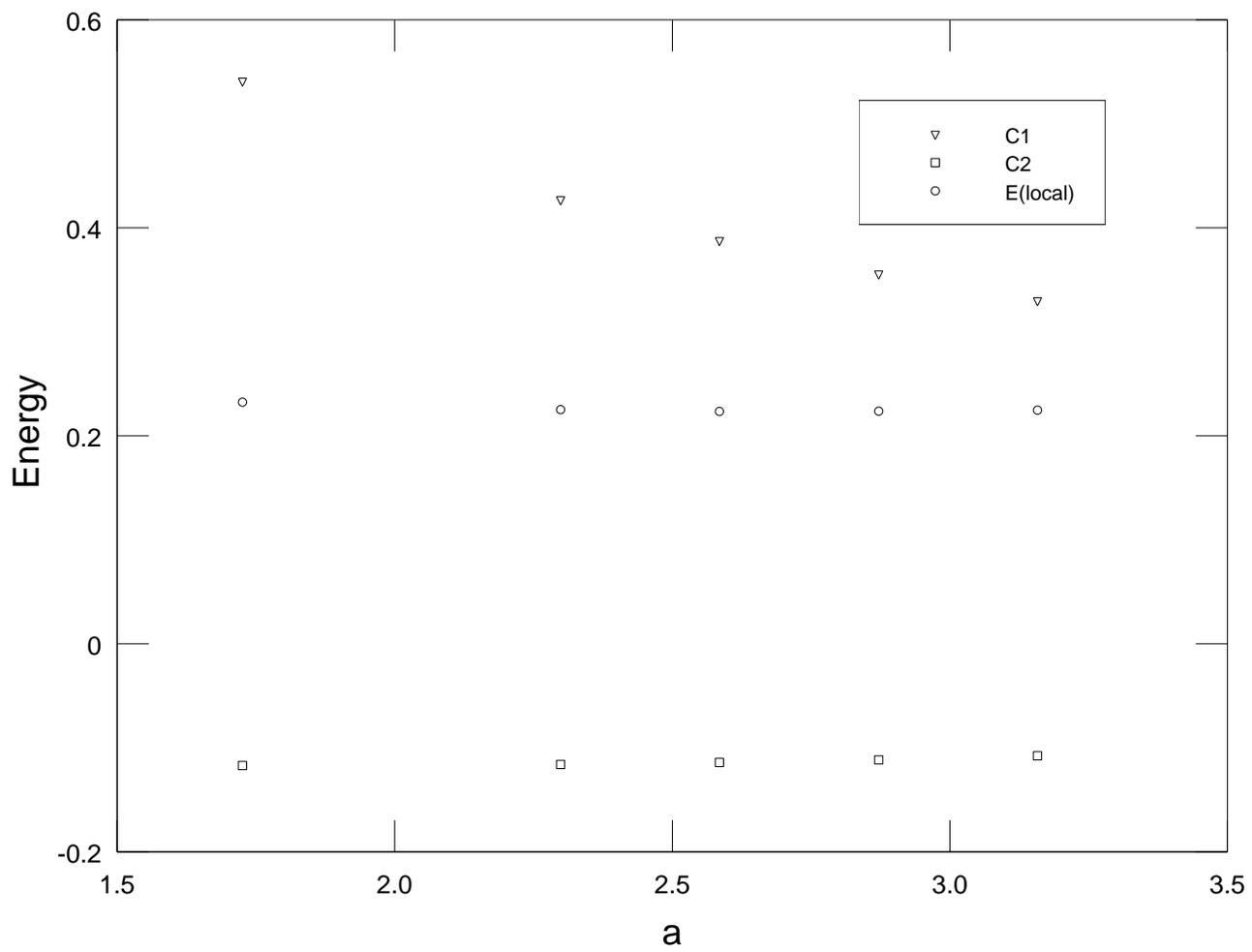
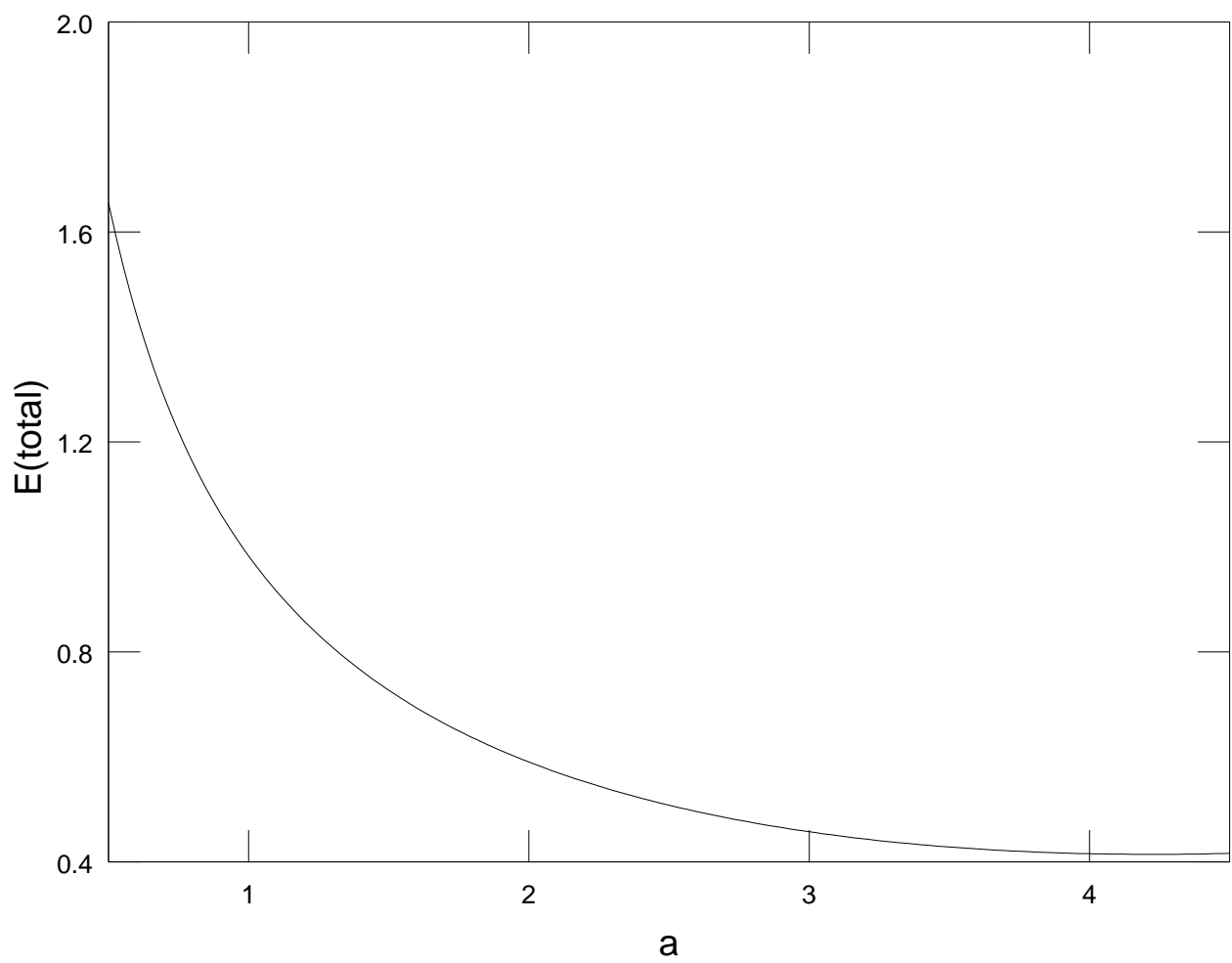
\end{document}